\documentclass[aip,numerical,reprint]{revtex4-1}
\usepackage{color}
\usepackage{graphicx}
\usepackage[squaren]{SIunits}
\usepackage[colorlinks=true,citecolor=blue,linkcolor=blue,pdfstartview=FitH,pdfauthor=Gross]{hyperref}

\begin{document}
\title[A Superconducting 180$\degree$ Hybrid Ring Coupler for circuit Quantum Electrodynamics]
      {A Superconducting 180$\degree$ Hybrid Ring Coupler for circuit Quantum Electrodynamics}

\author{E.~Hoffmann}\email{Elisabeth.Hoffmann@wmi.badw-muenchen.de}
\affiliation{Walther-Mei{\ss}ner-Institut, Bayerische Akademie der Wissenschaften, 85748 Garching, Germany}
\affiliation{Physik-Department, Technische Universit\"{a}t M\"{u}nchen, 85748 Garching, Germany}
\author{F.~Deppe}
\affiliation{Walther-Mei{\ss}ner-Institut, Bayerische Akademie der Wissenschaften, 85748 Garching, Germany}
\affiliation{Physik-Department, Technische Universit\"{a}t M\"{u}nchen, 85748 Garching, Germany}
\author{T.~Niemczyk}
\affiliation{Walther-Mei{\ss}ner-Institut, Bayerische Akademie der Wissenschaften, 85748 Garching, Germany}
\affiliation{Physik-Department, Technische Universit\"{a}t M\"{u}nchen, 85748 Garching, Germany}
\author{T.~Wirth}
\affiliation{Karlsruher Institut f\"{u}r Technologie (KIT), Physikalisches Institut, 76128 Karlsruhe, Germany}
\author{E.~P.~Menzel}
\affiliation{Walther-Mei{\ss}ner-Institut, Bayerische Akademie der Wissenschaften, 85748 Garching, Germany}
\affiliation{Physik-Department, Technische Universit\"{a}t M\"{u}nchen, 85748 Garching, Germany}
\author{G.~Wild}
\affiliation{Walther-Mei{\ss}ner-Institut, Bayerische Akademie der Wissenschaften, 85748 Garching, Germany}
\affiliation{Physik-Department, Technische Universit\"{a}t M\"{u}nchen, 85748 Garching, Germany}
\author{H.~Huebl}
\affiliation{Walther-Mei{\ss}ner-Institut, Bayerische Akademie der Wissenschaften, 85748 Garching, Germany}
\affiliation{Physik-Department, Technische Universit\"{a}t M\"{u}nchen, 85748 Garching, Germany}
\author{M.~Mariantoni}
\affiliation{Walther-Mei{\ss}ner-Institut, Bayerische Akademie der Wissenschaften, 85748 Garching, Germany}
\affiliation{Physik-Department, Technische Universit\"{a}t M\"{u}nchen, 85748 Garching, Germany}
\author{T.~Wei{\ss}l}
\affiliation{Walther-Mei{\ss}ner-Institut, Bayerische Akademie der Wissenschaften, 85748 Garching, Germany}
\affiliation{Physik-Department, Technische Universit\"{a}t M\"{u}nchen, 85748 Garching, Germany}
\author{A.~Lukashenko}
\affiliation{Karlsruher Institut f\"{u}r Technologie (KIT), Physikalisches Institut, 76128 Karlsruhe, Germany}
\author{A.~P.~Zhuravel}
\affiliation{B. I. Verkin Institute for Low Temperature Physics and Engineering, National Academy of Sciences of Ukraine, 61103 Kharkov, Ukraine}
\author{A.~V.~Ustinov}
\affiliation{Karlsruher Institut f\"{u}r Technologie (KIT), Physikalisches Institut, 76128 Karlsruhe, Germany}
\author{A.~Marx}
\affiliation{Walther-Mei{\ss}ner-Institut, Bayerische Akademie der Wissenschaften, 85748 Garching, Germany}
\author{R.~Gross}\email{Rudolf.Gross@wmi.badw-muenchen.de}
\affiliation{Walther-Mei{\ss}ner-Institut, Bayerische Akademie der Wissenschaften, 85748 Garching, Germany}
\affiliation{Physik-Department, Technische Universit\"{a}t M\"{u}nchen, 85748 Garching, Germany}

\date{\today}

\begin{abstract}
   Superconducting circuit quantum electrodynamics experiments with propagating microwaves require devices acting as beam splitters. Using niobium thin films on silicon and sapphire substrates, we fabricated superconducting 180\degree\ microstrip hybrid ring couplers, acting as beam splitters with center frequencies of about $6\,\giga\hertz$. For the magnitude of the coupling and isolation we find $-3.5\pm 0.5$\,\deci\bel\ and at least $-15\,\deci\bel$, respectively, in a bandwidth of 2\,\giga\hertz. We also investigate the effect of reflections at the superconductor-normal conductor contact by means of low temperature laser scanning microscopy. Our measurements show that our hybrid rings are well suited for on-chip applications in circuit quantum electrodynamics experiments.
\end{abstract}

\maketitle

In superconducting circuit quantum electrodynamics (QED)~\cite{Wallraff:2004a, Blais:2004, Chiorescu:2004}, intracavity microwave photons interact with solid-state artificial atoms~\cite{Makhlin:2001, Devoret:2004, Clarke:2008}. Both cavity and atom are realized by superconducting quantum circuits on a chip with characteristic frequencies in the microwave regime (1-10\,\giga\hertz). Recently, this field has been extended towards the study of propagating quantum microwaves. To this end, quantum optical techniques such as optical homodyne tomography~\cite{Leonhardt:1997} or photon-based quantum information processing and communication~\cite{Bouwmeester:2000, obrien_optical_2007} are being adapted to the microwave regime. One key element for the transformation from the optical to the microwave regime is the implementation of a beam splitter which is understood on the quantum level~\cite{Mariantoni:2005}. This allows the use of signal recovery techniques employing two amplifier chains and eliminating the (not yet available) single microwave photon detectors~\cite{bozyigit_measurements_2010}. Hereby, photon correlations can be accessed and all quadrature moments of propagating quantum microwaves and, simultaneously, those of the detector noise can be extracted~\cite{menzel_dual-path_2010}. The very same idea was recently used to characterize the black body radiation emitted by a 50\,\ohm\ load resistor~\cite{mariantoni_planck_2010}. Ideally, in experiments with propagating quantum microwaves a beam splitter has to be lossless. A device matching these conditions is the 180\degree\ hybrid ring which is entirely based on interference effects. Usually, microwave beam splitters are realized as normal conductive devices. However, for superconducting circuit QED the on-chip implementation of the beam splitter and the superconducting quantum devices under investigation would be favorable, avoiding reflections between various circuit parts and minimizing interconnect losses.

In this letter, we present a detailed study on low-loss superconducting hybrid rings fabricated from niobium microstrip lines on both silicon and sapphire substrates. For the magnitude of the coupling and isolation we find $-3.5\pm0.5\,\deci\bel$ and better than $-15\,\deci\bel$ in a bandwidth of up to 2\,\giga\hertz, respectively. We note that the isolation increases when reducing the bandwidth, reaching a maximum value of better than $-60\,\deci\bel$ at the center frequency. Our measurements indicate that the device performance is limited by reflections between the superconducting parts on the chip and the normal conducting microwave connectors. This conclusion is based on our data obtained by low temperature laser scanning microscopy (LTLSM)~\cite{Zhuravel:2006, zhuravel_measurement_2006}. Our experiments indicate that our hybrid ring couplers are highly suitable for integration into superconducting circuit QED experiments~\cite{Maya:2003}, ultimately allowing for studies of propagating quantum microwaves~\cite{bozyigit_measurements_2010, menzel_dual-path_2010, mariantoni_planck_2010} and applications in quantum information processing.

\begin{figure}[tb]
  \centering
   \includegraphics[width=\columnwidth]{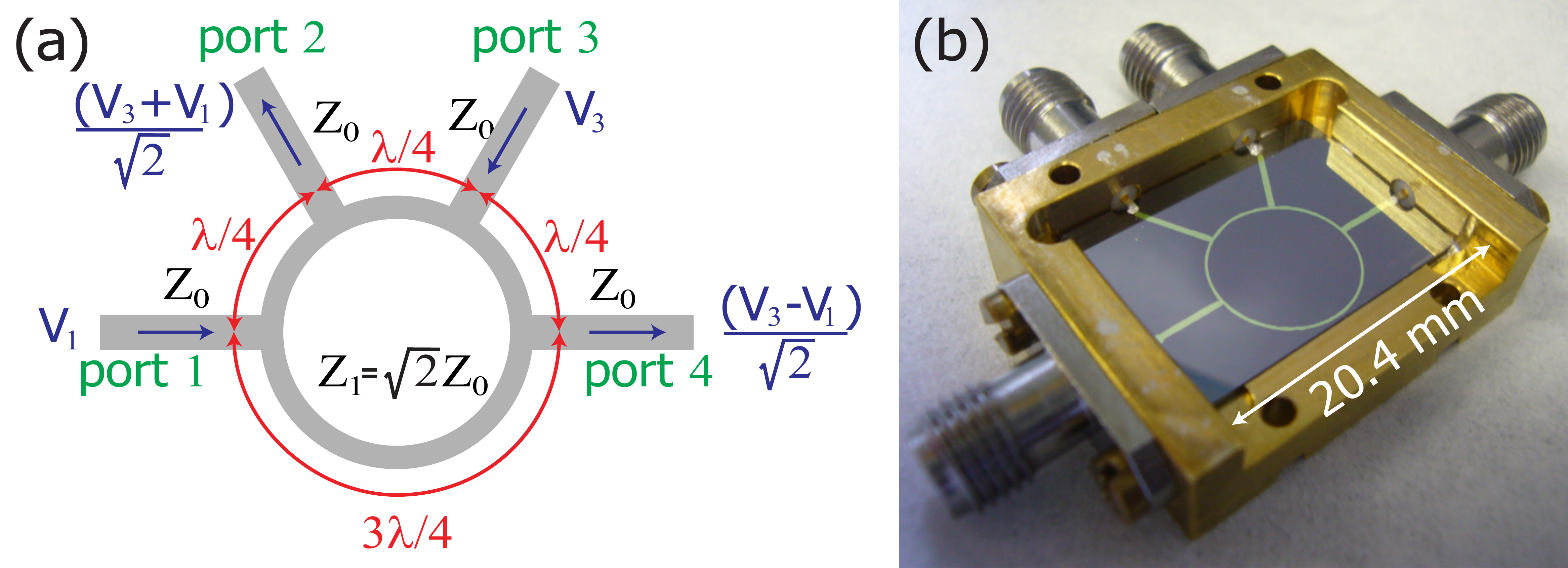}
   \caption{(a) Schematic of a 180\degree\ hybrid ring. In this configuration port one and three act as input ports while port two and four are output ports. (b) Photograph of a niobium hybrid ring fabricated on a silicon substrate and mounted inside a gold-plated copper box.}
   \label{Pic:Hoffmann_Sample_2010}
\end{figure}

The 180\degree\ hybrid ring is sketched in Fig.~\ref{Pic:Hoffmann_Sample_2010}(a). It consists of a superconducting ring with four signal ports. The circumference $U=1.5\lambda$ of the ring
determines the center frequency $f_0=v_\mathrm{ph}/\lambda$ of the device. Here, $v_\mathrm{ph}$ is the phase velocity of electromagnetic waves and $\lambda$ the wavelength. An input signal of frequency $f$ incident at port one (or three) is split into its clockwise and counterclockwise propagating components which interfere constructively ($3\,\deci\bel$ coupling) at ports two and four, whereas they interfere destructively (isolation) at ports three and one. When two signals are applied to port one and port three, their sum and difference is present at port two and four, respectively. To avoid reflections and to guarantee an equal splitting of the signal, the impedance of the ring~\cite{Pozar:2005, Collin:2000} must be chosen as $Z_1{=}Z_0\sqrt{2}{=}71\,\ohm$ for a feed line impedance $Z_0{=}50\,\ohm$.

The hybrid rings are based on 200\,nm thick niobium films deposited by magnetron sputtering and patterned by optical lithography and reactive ion etching using SF$_6$. Niobium is chosen due to its high critical temperature of 9.2\,K. As substrate materials we use silicon (thickness 525\,\micro\meter, dielectric constant $\varepsilon_\mathrm{r}{=}11.9$) covered by 50\,nm of silicon dioxide, as well as sapphire (thickness 500\,\micro\meter, $\varepsilon_\mathrm{r}{=}10.2$).
Although recent measurements at millikelvin temperatures~\cite{OConnell:2008} show loss tangents of ${\simeq}\,10^{-5}$ for both crystalline materials sapphire and silicon, we need to verify to what extent the amorphous SiO$_2$ coating of our silicon substrates affects the device performance. The radius of all studied hybrid rings is $4.78\,\milli\meter$, corresponding to $f_0=5.6\,\giga\hertz$ ($6\,\giga\hertz$) for the samples on silicon (sapphire). The microstrip lines forming the input and output ports are $420\,\micro\meter$ (490\,\micro\meter) wide for the devices on silicon (sapphire). The width of the strip forming the ring is $171\,\micro\meter$ (221\,\micro\meter). For the characterization of the microwave properties, the chip is mounted inside a gold-plated copper box as shown in Fig.~\ref{Pic:Hoffmann_Sample_2010}(b) and then cooled down in a $^4$He-cryostat. We recorded the coupling and isolation properties from seven (four) different hybrid rings fabricated on silicon (sapphire) substrates. Each chip is remounted, cooled down, and remeasured several times to test the reproducibility.

\begin{figure}[tb]
  \centering
   \includegraphics[width=\columnwidth]{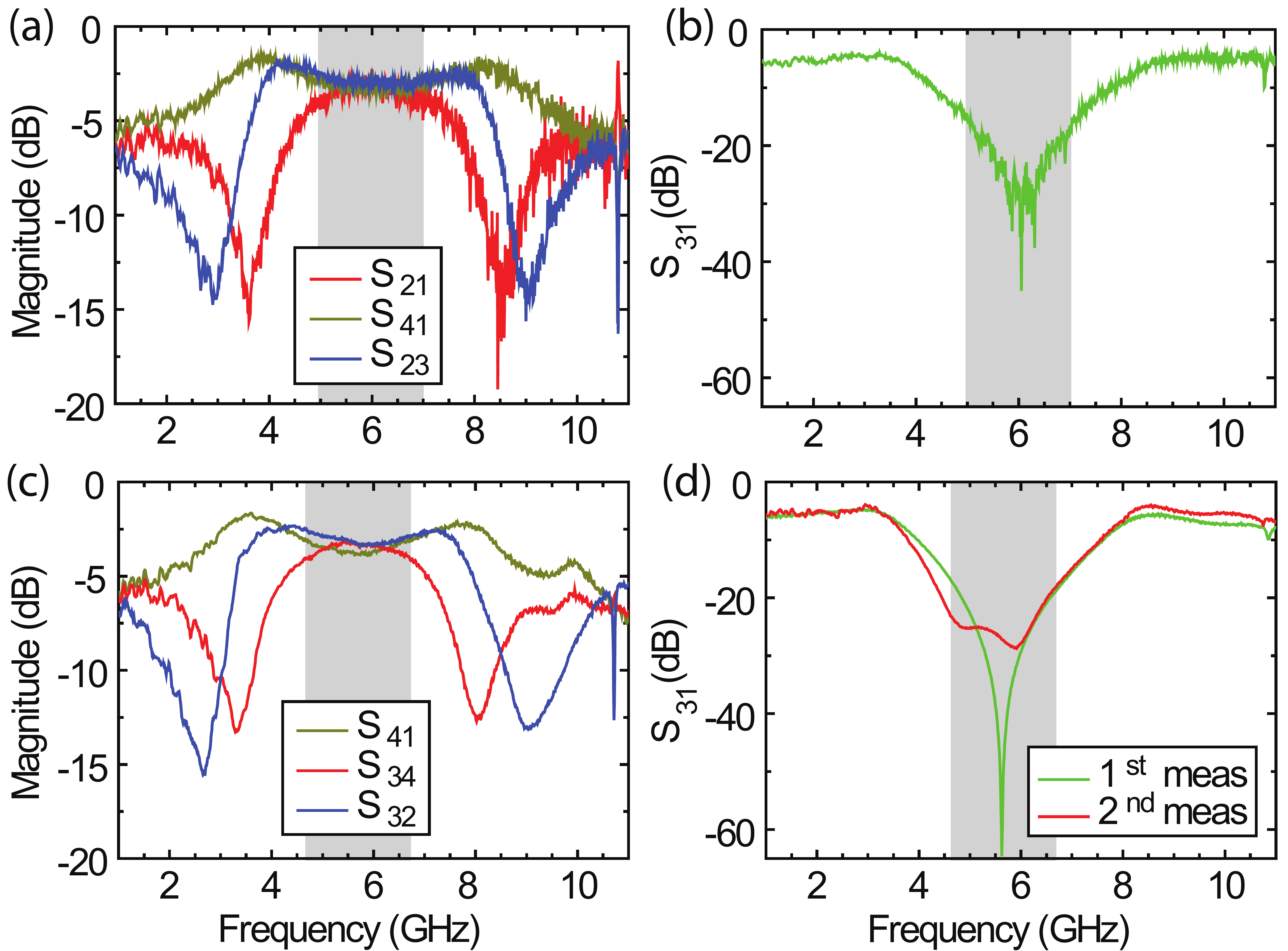}
   \caption{Typical transmission data for hybrid rings on sapphire (panel a and b) and silicon (panel c and d) substrates measured at $4.2\,\kelvin$ with an input power of $-40\,\deci\bel\milli$. For both materials we observe a coupling of $-3.5\pm0.5\,\deci\bel$ and an isolation of at least $-15\,\deci\bel$ within a bandwidth of $2\,\giga\hertz$ (grey background) around the center frequency. Note that the different noise level is not related to the substrate material but is caused by difference in the measurement protocol: no averaging in (a) and (b); 20 times averaging in (c) and (d). Additionally, in (d) $S_{31}$ is shown for the same sample after remounting it. While the green curve shows a close-to-ideal spectrum, the red curve shows a ``hump" caused by reflections at the chip-connector contact (cf. Fig.~\ref{Hoffmann_Erlangen}). In the coupling spectra of (c), such reflections manifest themselves as an asymmetry. In all measurements a box resonance at approximate 11\,\giga\hertz\ can be seen.}
   \label{Hoffmann_Data}
\end{figure}

The performance of the hybrid rings is studied by measuring the scattering matrix $S_{ij}$ ($i,j=1\dots4$) using a two-port vector network analyzer. We only measured the scattering parameters with $i\ne j$ by connecting ports $i$ and $j$ to the network analyzer, while the other two ports are terminated right at the sample box with $50\,\ohm$ loads. The characteristics of our hybrid rings fabricated on a sapphire and silicon substrates are shown in Figs.~\ref{Hoffmann_Data}(a)-(d). Figure~\ref{Hoffmann_Data}(a) and Fig.~\ref{Hoffmann_Data}(c) display the $S$-parameters for constructive interference at the respective output port. For both substrate materials we find a coupling magnitude of $-3.3\pm0.2$\,\deci\bel\ at the center frequency $f_0$ as expected for a $-3\,\deci\bel$ beam splitter. Within a bandwidth of 2\,\giga\hertz\ around $f_0$, the coupling magnitude is in the range of $-3.5\pm0.5$\,\deci\bel. Well outside this frequency window it drops below $-10\,\deci\bel$, reflecting the considerable mismatch between the device circumference $U$ and $1.5\lambda$ at the test frequency. The isolation of the devices is shown in Fig.~\ref{Hoffmann_Data}(b) and Fig.~\ref{Hoffmann_Data}(d) for sapphire and silicon substrates, respectively. In both cases the isolation magnitude exceeds $-15\,\deci\bel$ within the full bandwidth of 2\,\giga\hertz\ around $f_0$, showing the excellent performance of the hybrid rings. We note that despite the SiO$_2$ coating of the silicon substrate, the performance of our hybrid rings is robust with respect to dielectric losses.

In some of our devices we find characteristic changes in the transmission data when remounting and remeasuring the same device. First, the frequency of the maximum isolation may shift and a ``hump" may appear in the spectrum as shown in Fig.~\ref{Hoffmann_Data}(d). Concurrently, the coupling spectra may become asymmetric and their magnitudes at the center frequency may vary slightly as indicated in Fig.~\ref{Hoffmann_Data}(c). The likely origin of these features are reflections at the contact between the superconducting on-chip feed lines and the normal conducting microwave connectors, which affect the interference pattern in the ring. To explore these artifacts, we visualize the effects of reflections by measuring the response of a hybrid ring on a silicon substrate to local heating by a focused laser beam. This method is known as low temperature laser scanning microscopy~\cite{Zhuravel:2006}. In these experiments, a focused laser beam is scanned across the chip surface and the change in the transmission parameter $S_{42}$ is recorded as function of the beam position. Local heating by the focused laser beam results in quasi-particle generation in the niobium film~\cite{Koelle:1994}. Therefore, local ohmic dissipation proportional to the high frequency electrical field changes the transmission magnitude.

\begin{figure}[tb]
  \centering
   \includegraphics[width=\columnwidth]{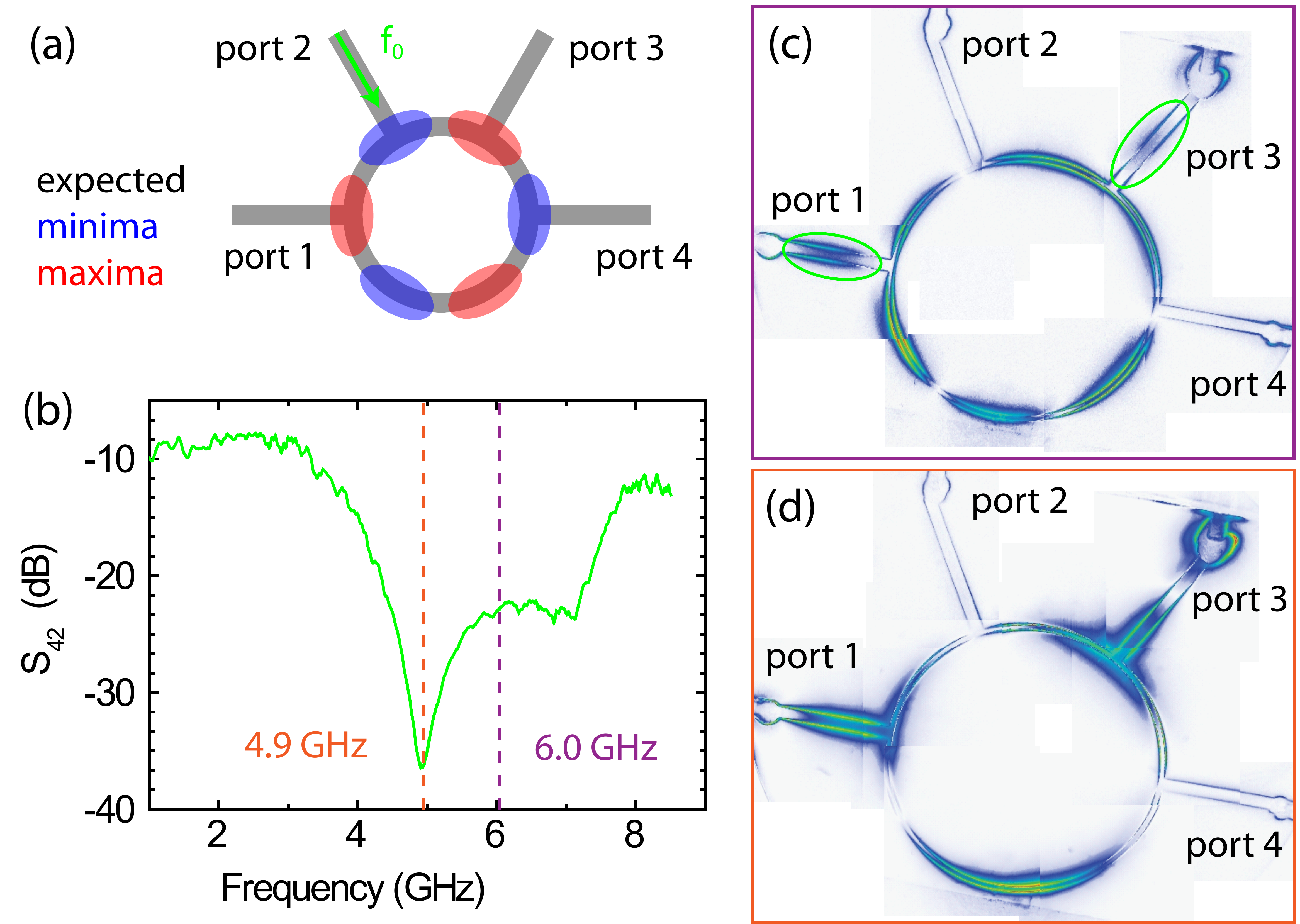}
   \caption{(a) Expected spatially resolved transmission magnitude at the center frequency $f_0$ where the isolation should reach its minimum. Port two is used as input port. (b) Measurement of the integral $S_{42}$-isolation at $4.2\,\kelvin$ with a network vector analyzer attached to the LTLSM apparatus.  (c) LTLSM data taken at the ``hump" (6\,\giga\hertz; color code: transmission magnitude). (d) LTLSM data taken at $4.93\,\giga\hertz$, where the isolation is maximal.}
   \label{Hoffmann_Erlangen}
\end{figure}

The LTSLM images shown in Fig.~\ref{Hoffmann_Erlangen}(c) and Fig.~\ref{Hoffmann_Erlangen}(d) can be understood in a straightforward way. With an input signal at the center frequency $f_0$ at port two, the interference in the ring is expected to lead to three maxima, two of them located at port one and three, and three minima, two of them are expected to be found at port two and port four, as depicted in Fig.~\ref{Hoffmann_Erlangen}(a). The transmission $S_{42}$ without laser irradiation measured with a vector network analyzer in the LTLSM setup is plotted in Fig.~\ref{Hoffmann_Erlangen}(b) showing the combination of a ``hump" and a shifted minimum at $4.93\,\giga\hertz$. In Fig.~\ref{Hoffmann_Erlangen}(c), the spatially resolved transmission magnitude at the ``hump"-frequency is displayed. First, we notice maxima in the feed lines (marked green), which indicate standing waves caused by reflections at the connectors. Second, although the electric field maxima and minima on the ring are slightly shifted from their ideal positions, they are still located near the expected ports. Consequently, there remains significant isolation between ports two and four. Third, when changing the excitation frequency to the isolation maximum at 4.93\,\giga\hertz, the ideal interference pattern is restored in the ring as shown in Fig.~\ref{Hoffmann_Erlangen}(d). Nevertheless, there is still a significant signal in the feed lines due to the reflections from the connectors.

In conclusion, we have fabricated superconducting 180\degree\ hybrid ring couplers on both sapphire and silicon substrates. Within a 2\,\giga\hertz\ bandwidth around the center frequency of around 6\,\giga\hertz, the devices show an almost ideal coupling of $-3.5\pm0.5\,\deci\bel$. Furthermore, we find an isolation of at least $-15\,\deci\bel$. The observed imperfections are clearly attributed to remaining reflections at the transition between the superconducting on-chip feed lines and the normal conducting microwave connectors, demonstrating the importance of proper mounting. The performance of our hybrid rings is suitable for further experiments with propagating quantum microwaves, e.g., in the spirit of those presented in Ref.~\onlinecite{menzel_dual-path_2010} or Ref.~\onlinecite{mariantoni_planck_2010}.

\acknowledgments{We thank A. Emmert for useful discussions. For financial support, we acknowledge the Deutsche Forschungsgemeinschaft via SFB~631 and CFN, EU project SOLID and the German Excellence Initiative via NIM.}

\end{document}